\def\be{\begin{equation}}
\def\ee{\end{equation}}
\def\ber{\begin{eqnarray}}
\def\eer{\end{eqnarray}}
\def\bers{\begin{eqnarray*}}
\def\eers{\end{eqnarray*}}

\documentclass[aps,prapplied,twocolumn,showpacs,superscriptaddress, amsmath,amssymb,reprint,longbibliography]{revtex4-2}

\usepackage{dcolumn}   
\usepackage{amsmath}
\usepackage{multirow}
\usepackage{graphicx}    
\usepackage{subfigure}
\usepackage{comment}
\usepackage{color}
\usepackage{makecell}
\usepackage{longtable}
\usepackage{natbib}
\usepackage{ifthen}
\usepackage{nicefrac}
\newboolean{includefigs}
\setboolean{includefigs}{true}      
\newboolean{includetext}
\setboolean{includetext}{true}     
\newcommand{\condcomment}[2]{\ifthenelse{#1}{#2}{}}

\begin{document}

\title{Ternary Alkali Metal Copper Chalcogenides ACuX (A= Na, K and X= S, Se, Te): Promising Candidate for Solar Harvesting Applications}

\author{Gurudayal Behera}
\affiliation{Department of Energy Science and Engineering, IIT Bombay, Powai, Mumbai 400076, India}

\author{Surabhi Suresh Nair}
\affiliation{Department of Physics, Khalifa University of Science and Technology, Abu Dhabi 127788, United Arab Emirates}

\author{Nirpendra Singh}
\affiliation{Department of Physics, Khalifa University of Science and Technology, Abu Dhabi 127788, United Arab Emirates}

\author{K. R. Balasubramaniam}
\email{bala.ramanathan@iitb.ac.in}
\affiliation{Department of Energy Science and Engineering, IIT Bombay, Powai, Mumbai 400076, India}

\author{Aftab Alam}
\email{aftab@iitb.ac.in}
\affiliation{Department of Physics, Indian Institute of Technology, Bombay, Powai, Mumbai 400076, India}

\begin{abstract}
We report a comprehensive first--principles study of the relative stability of the various possible crystal structures, and the electronic and optical properties of ternary alkali metal chalcogenides ACuX (A= Na/K and X= S/Se/Te) compounds through density functional theory (DFT) calculations. The energetics and phonon spectra of greater than 700 structures were compared, and seven possible stabilized structures of six ACuX compounds were identified using the fixed composition evolutionary search method. Our electronic band structure simulation confirms that all the ternary ACuX compounds are direct band gap semiconductors, with the band gap lying between 0.83 eV to 2.88 eV. These compounds exhibit directly allowed electronic transitions from the valence band to the conduction band, which leads to a significant strength of optical transition probability. This yields a sharp rise in the optical absorption spectra (ranging between 10$^4$ to 10$^5$ cm$^{-1}$) near the energy gap. The estimated spectroscopic limited maximum efficiency (SLME) is about 18\% for an 8 $\mu$m thick NaCuTe film. For other ACuX compounds, the SLME ranges between 10\% to 13\%. In addition, we also explored the feasibility of these ternary ACuX compounds for photocatalytic water splitting applications and found that they can be promising candidates as photocathodes for hydrogen evolution reactions. With a large spread in the band gap and interesting band topology near Fermi level, these chalcogenides can be quite fertile for other energy applications such as thermoelectric, LED, \emph{etc}. 
\end{abstract}
\date{\today}
\maketitle
\section{Introduction}
In recent years, multinary chalcogenide semiconductors have witnessed extensive applications in the development of thin-film solar photovoltaic (PV) devices. For example, CuIn$_\mathrm{1-x}$(Se,Se)$_\mathrm{2}$ (CIGS), Cu$_\mathrm{2}$ZnSnS$_\mathrm{4}$, CdTe, SnS, Sb$_\mathrm{2}$S$_\mathrm{3}$, \emph{etc.} have shown immense promise in thin film solar cells \cite{saparov2022next,palchoudhury2020multinary,shin2017defect,sharma2022review}. Among them, CdTe and CIGS are most prevalent due to their direct band gaps and high light absorption capabilities, enabling them to achieve notable solar cell efficiencies\cite{polman2016photovoltaic,martin2008solar,ouendadji2011theoretical,eyderman2014near}.  It is however also known that CdTe and CIGS come with certain drawbacks. They contain toxic elements, namely Cd and Te, as well as elements that are relatively scarce in the earth's crust, such as Te, In, and Ga, which is a major concern that could limit market feasibility for sustainable, cost-effective, and large-scale manufacturing. Thus, identifying efficient light absorber materials in thin film-based PV devices with band gaps in the visible range is a crucial challenge. Apart from this, there are few other major issues in existing solar absorbers employed in PV devices are \emph{e.g.,} organic moieties that cause degradation over short carrier lifetimes (\emph{i.e.,} organic/inorganic halide perovskites \cite{eyderman2014near,stranks2013electron,liu1999electronic,sha2015efficiency}), and the presence of defect states causing high recombination rate \cite{stranks2014recombination,tress2016inverted}. As such, finding alternative materials is an urgent need of the hour. The material should acquire few prerequisite criteria such as low cost (abundant), high carrier lifetime, high absorption and defect tolerance. 

Recently, alkali metal ternary copper telluride ACuTe (A=Na, K) has been proposed to be efficient light absorber for thin film solar cells, satisfying most of the above criteria\cite{dahliah2021high}.  They are reported to be direct band gap semiconductors having band gap values of 1.43 eV and 1.63 eV for NaCuTe and KCuTe, respectively. The high absorption coefficient ($\sim$ 10$^4$ cm$^{-1}$) in the visible range and a reasonably high carrier lifetime due to low deep level defects suggest their potentiality as efficient solar absorbers. The high abundance of constituent elements is an added advantage. This has motivated us to further investigate the feasibility of other chalcogenides (S and Se) based ternary compounds as solar absorbers.

The ternary alkali metal copper chalcogenides ACuX (A=Na, K and X=S, Se, Te) have been experimentally synthesized by Savelsberg and Schfer in 1978 \cite{savelsberg1978ternare,savelsberg1978darstellung}. They found that KCuS crystallizes in orthorhombic structure (space group Pna2$_1$), while KCuSe/Te has a hexagonal structure with space group P6$_3$/mmc. Similarly, NaCuSe/Te is reported to crystallize in tetragonal structures with space group P4/nmm. Since then, no further experimental as well as theoretical studies have been reported on this class of compounds. Recently, Vaitheeswaran {\it et al.} \cite{parveen2018exploring} studied the structural, electronic, and optical properties of ternary KCuSe and KCuTe using {\emph ab-initio} calculations. They reported these compounds to be semiconductors with suitable band gaps that could be promising as solar absorbers in PV, photodetectors, and other optoelectronic device applications. On a similar line, Boualleg \emph{et al.}\cite{boualleg2022ab} reported the phase transition and thermal properties of KCuSe and KCuTe. They reported these two compounds to stabilize in orthorhombic and tetragonal phases, respectively in contrast to the experimental hexagonal structure. The dynamical stability study also confirmed the existence of other possible crystal structures for KCuSe and KCuTe compounds. For other ternary chalcogenides \emph{i.e.,} NaCuSe, NaCuTe, and KCuS, no further studies (both experiment and theory) are reported reassessing their crystal structures. NaCuS is never reported earlier and hence its crystal structure is not known. The conflict between experimental and theoretically predicted crystal structures and the prediction of the most stable structural phases needs to be addressed. In addition, a thorough study to investigate the potential of all these ternary compounds for various solar harnessing applications can be extremely useful.

In this work, we aim to present a detailed insight into the structural stability, electronic, and optical properties of ternary ACuX compounds assessing their potential as efficient light absorber using first--principles Density Functional Theory (DFT)  calculations. In particular, we utilized a fixed composition evolutionary search method to accurately predict the correct crystal structure of all these six compounds. The method generates greater than 700 structures (depending on the system) and scrutinize the most stable out of them based on certain factors. This has been a powerful method to predict ground state structures for a given system. Interestingly, though the crystal structure of a few systems matches with those predicted experimentally, there are other systems where the most stable structure is few meV lower in energy as compared to the experimental ones. Lattice dynamics calculations are also performed in parallel to evaluate the dynamically stable structures for each compound.
We further simulated the electronic and optical properties of all six ACuX (A=Na/K, X=S/Se/Te) compounds in their most stable phases. All six ACuX compounds are found to be direct band gap semiconductors with band gap values ranging between 0.83 eV to 2.88 eV. These band gaps are simulated using the most accurate hybrid exchange-correlation functional. The optical simulation confirms the anisotropic nature of the absorption spectra having different in-plane and out-of-plane values, with absorption coefficient $\sim$10$^4$ cm$^{-1}$ in the visible range. The theoretically predicted spectroscopic limited maximum efficiency (SLME) is found to lie in the range 10 \% to 18 \% for a thickness of 8 to 10 $\mu$m under 1 sun illumination AM1.5G. Further, we evaluated the potential of all stable ACuX compounds as photocathodes for water splitting applications based on their band edge positions with respect to water redox potential. 

\section{Computational Details}
The fixed composition evolutionary search method is utilized for predicting the crystal structure, as employed in the USPEX package\cite{oganov2011evolutionary,lyakhov2013new,oganov2006crystal}.  An initial population of 20 structures is generated randomly\cite{lyakhov2013new}, while the subsequent generations are produced due to variation operators, which include heredity (50\%), random symmetric structure generation (30\%), and soft mutation (20\%). Subsequently, a comprehensive exploration is carried out by generating more than 700 crystal structures depending on the six compounds. Our enthalpy calculations based on DFT identify optimal structures from each generation within the selection criterion. The procedure for local optimization is executed in a five-step process utilizing the generalized gradient approximation (GGA)\cite{perdew1997generalized} within the Perdew-Burke-Ernzerhof functional (PBE)\cite{perdew1996generalized}, implemented in the Vienna Ab--initio Simulation Package (VASP) \cite{vasp,kresse1993ab,kresse1996efficient}. For a more accurate calculation of the electronic structure (band gap), the screened hybrid HSE06 functional \cite{heyd2003hybrid,heyd2004efficient} was employed. The kinetic energy cut--off for the plane wave basis set was set to 520 eV. The Brillouin zone (BZ) integration was done using a $\Gamma$--centered scheme with 12$\times$12$\times$6 k--meshes for ionic relaxations and 16$\times$16$\times$8 for self-consistent-field calculations. The density of states (DOS) was calculated using the tetrahedron method with Bl{\"o}ch corrections\cite{bolch}. Electronic band structure within the HSE06 was simulated using 8$\times$8$\times$4 k--mesh. All the atoms in the unit cell are fully relaxed using the conjugate gradient method until the force (energy) converges below 0.001 eV/\AA\ (10$^{-7}$ eV). The phonon spectra are calculated using the supercell approach as implemented in the phonopy package \cite{phonopy,parlinski1997first}. Effect of spin--orbit coupling (SOC) was included for all the compounds. 
Formation energy ($\Delta E_\mathrm{F}$) of ACuX (A; Na/K and X; S/Se/Te) were calculated using the following expression,

\begin{equation}
\Delta E_F=\frac{E_{Tot}-[nE_A-lE_{Cu}-mE_X]}{N}
\end{equation}
where “$E_{Tot}$” is the total energy of the target compound, “$n$”, “$l$”, and “$m$” are the numbers of Na/K, Cu, and S/Se/Te atoms present in the compound, while $N$ is the total number of atoms present in the cell. E$_A$, E$_{Cu}$, and E$_X$ are the total energies per atom of the constituent elements in their respective bulk equilibrium structures.

To simulate the optical properties, the frequency--dependent dielectric constants were calculated within the independent particle approximation (IPA) \cite{gajdovs2006linear,adler1962quantum,wiser1963dielectric}, as implemented in VASP. To measure the theoretical maximum possible efficiency, we have used an improved version of the Shockley-Queisser (SQ) efficiency limit known as spectroscopic limited maximum efficiency (SLME) proposed by Yu \emph{et al.} \cite{yu2012identification} using SL3ME code\cite{bercx2018exceeding}. More details about the theoretical formulation of optical calculations and  SLME can be found in Sec I of the supplementary information (SI)\cite{supplment}.

\section{Results and Discussion}

\subsection{Energetics and Dynamical Stability of ACuX}

\begin{table}[t!]%
	\centering
	\caption{\label{tab:table1}%
		Formation energies ($\Delta E_F$) of experimentally reported and theoretically simulated lowest energy structures for six ACuX compounds. `*' indicates the experimentally reported space groups. $\Delta$ is the energy difference between the $\Delta E_F$ of experimental and theoretical structures. }
	\begin{ruledtabular}
	\begin{tabular}{cccc}
		\bf{Compounds}	&\bf{Space}&\bf{$\Delta E_F$}& \bf{$\Delta$}\\
		&{\bf Group} &\bf{(meV/atom)}&\bf{(meV/atom)}\\ 
		\hline \\
		NaCuS&Pna2$_1$&-708&Not synthesized\\ \\
		
		NaCuSe&P4/nmm*\cite{savelsberg1978ternare}&-617&20\\
		&P6$_3$/mmc&-637&\\ \\
		
		NaCuTe&P4/nmm*\cite{savelsberg1978ternare}&-493&17\\
		&P6$_3$/mmc&-510&\\ \\
		
		KCuS&Pna2$_1$*\cite{savelsberg1978darstellung}&-768&\\ \\
	
		KCuSe&P6$_3$/mmc*&-687&12\\ 
		&Pnma\cite{savelsberg1978ternare}&-695&\\ \\
		
		KCuTe&P6$_3$/mmc*\cite{savelsberg1978ternare}&-595&\\
	
	\end{tabular}
\end{ruledtabular}
\end{table}
Out of all the structures (more than 700 structures) generated by USPEX package \cite{oganov2011evolutionary,lyakhov2013new,oganov2006crystal}, we selected nine structures with the lowest formation energies for each ternary ACuX (A= Na, K and X=S, Se, and Te) compounds. The space groups (SG) of these nine structures are Pnma, Pna2$_1$, P6$_3$/mmc, P4/nmm, Cmcm, P2$_1$/m, F$\bar{4}$3m, P6$_3$mc, and C2/c, respectively. The prototype crystal structures of such space groups are shown in Fig.~S1 of the SI \cite{supplment} and their optimized lattice parameters and corresponding formation energies obtained from DFT calculations for all the six different ACuX compounds are listed in Table~S1 to S6 in SI \cite{supplment}. Table~\ref{tab:table1} display the space group and formation energies ($\Delta E_F$) of the experimentally reported crystal structures and the energetically most stable simulated crystal structures and their energy difference ($\Delta$). It is observed from Table~\ref{tab:table1} that NaCuS (never studied before) crystalizes in the Pna2$_1$ space group. However, energetically NaCu(Se/Te) should stabilize in hexagonal structure (SG: P6$_3$/mmc), whose $\Delta E_F$ is $\sim$17-20 meV/atom lower than those of experimentally reported tetragonal structure (space group P4/nmm). KCuS was found to stabilize in the orthorhombic structure having space groups Pna2$_1$, which agrees with the experimental prediction\cite{savelsberg1978darstellung}. In contrast, KCuSe is found to stabilize in the orthorhombic crystal structure, which is 12 meV/atom lower in energy than the experimentally reported hexagonal structure\cite{savelsberg1978ternare}. KCuTe stabilizes in the hexagonal structure (SG:P6$_\mathrm{3}$/mmc), which matches with the experimentally findings\cite{savelsberg1978ternare}.

To further assess the phase stability, specially for NaCuSe, NaCuTe and KCuSe compounds where the experimental structure is higher in energy (by 12 to 20 meV/atom) as compared to the theoretically optimized one, we simulated their phonon dispersion which will help to evaluate the dynamical stability. The phonon dispersion for all the experimental and theoretically optimized ACuX compounds in different space groups (as shown in Table~\ref{tab:table1}) are shown in Fig.~S2 of SI\cite{supplment}. For NaCuS (SG: Pna2$_1$), NaCuSe/Te (SG: P6$_3$/mmc), and all the KCuX compounds in their respective space groups, the phonon frequencies are found to be positive, confirming the dynamical stability for all these compounds. A very small imaginary frequency appears at/around the $\Gamma$ point in the orthorhombic structures of KCuSe (SG: Pnma), which may be due to the limited supercell size used in this calculation.
Interestingly, the phonon spectrum for the experimentally reported NaCuSe and NaCuTe (SG: P4/nmm) show appreciable imaginary frequencies (see Fig.~S2(b and d) of SI\cite{supplment}), indicating their instability in tetragonal phase. Most likely, they crystallize in hexagonal crystal structure, \emph{i.e.,} P6$_3$/mmc space groups, which is favored by both formation energy and phonon dispersion data.  We believe the structural characterization of these two systems should be revisited. In summary, based on the formation energies and phonon spectra, we have chosen seven most stable ternary ACuX compounds, \emph{i.e.,} orthorhombic NaCuS and KCuS (SG: Pna2$_1$), hexagonal NaCuSe/Te and KCuSe/Te (SG: P6$_3$/mmc) and the orthorhombic $\&$ hexagonal KCuSe (SG: Pnma) to further explore their optoelectronic properties.

\begin{figure*}[t]
	\centering
	\includegraphics[scale=0.5]{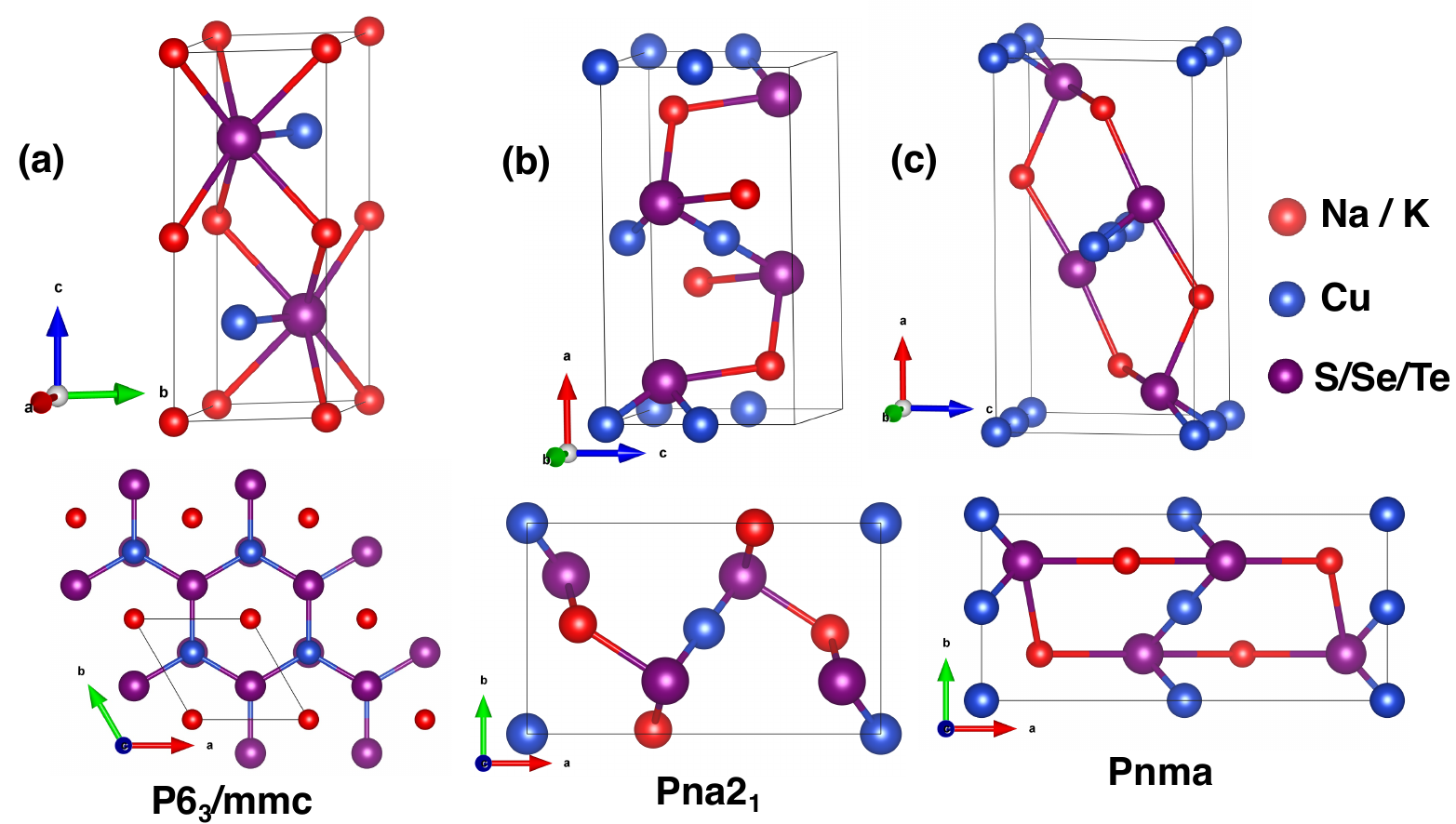}
	\caption{Prototype crystal structures of ACuX compounds in different space groups, as reported in Table \ref{tab:table1} (a) ACuX (X=Se, Te) (P6$_3$/mmc), (b) ACuS (Pna2$_1$), (c) KCuSe (Pnma), respectively. The set of top and bottom figures indicate the view of the respective structures from (110) and (001) orientations, respectively. }
	\label{fig1}
\end{figure*}

\begin{figure*}[t]
	\centering
	\includegraphics[scale=0.51]{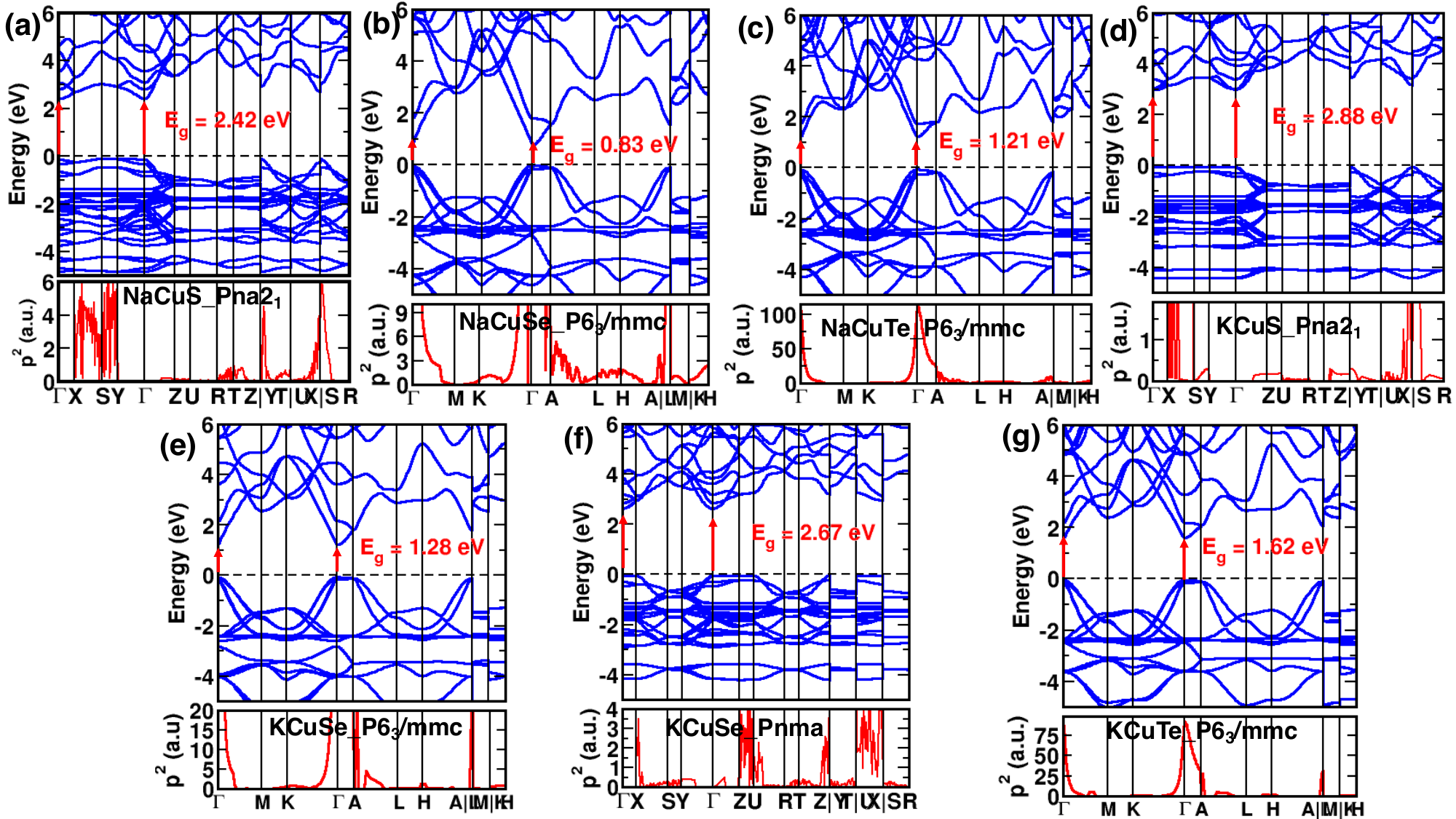}
	\caption{Electronic band structures (above panels) and optical transition probabilities (square of dipole transition matrix elements) (below panels) of ACuX compounds in their respective lowest energy structure using HSE06 functional. For KCuSe, results for both P6$_3$/mmc and Pnma are shown due to a relatively small energy difference between the two. Fermi level (E$_\mathrm{f}$) is set at 0 eV. }
	\label{fig2}
\end{figure*}

\begin{table*}[htbp]%
	\centering
	\caption{\label{tab:table2}%
		Theoretically optimized lattice parameters, bond lengths, and band gaps of energetically most stable ACuX (A=Na, K; X=S, Se, Te) compounds. Experimental results\cite{savelsberg1978ternare,savelsberg1978darstellung} are given wherever available.}
	\begin{ruledtabular}
 \resizebox{\textwidth}{!}{
  \small{
  \begin{tabular}{cccccc}
  			\bf{Compounds}	&\bf{Space group}&\bf{Lattice Constants}&\bf{Bond length}&\bf{Band gap (eV)}\\
			&& \bf{ (\AA)}&\bf{(\AA)}&\bf{ PBE/HSE06}\\
			\hline
			NaCuS&Pna2$_1$&a= 9.48, b= 5.65 
			&Cu--S= 2.17, &1.55/2.42\\
                &&c=5.13&Na--S= 2.89, 2.79&&\\
			\\
			NaCuSe&P6$_3$/mmc&a = 4.14, c= 8.34&Cu--Se= 2.39,&0/0.83\\
			&&& Na--Se= 3.17&&\\
                \\
			NaCuTe&P6$_3$/mmc&a= 4.42, c= 8.69&Cu--Te= 2.55,	
			&0.49/1.21\\
                &&&Na--Te= 3.35&&\\
			\\
			KCuS&Pna2$_1$&a= 10.88, b= 6.34, &Cu--S= 2.16,   &1.88/2.88\\ 
                &&c= 5.34&K--S= 3.21,3.19&\\
			&&a= 10.66, b= 6.20, &\\
                &&c= 5.32 (exp.\cite{savelsberg1978darstellung})&&\\
			\\
			KCuSe&Pnma&	a= 11.84, b= 5.43, & Cu--Se= 2.28, 	&1.72/2.67\\
                &&c= 6.36&K--Se=3.31, 3.39&&\\
			&P6$_3$/mmc&	a= 4.20, c= 9.92&Cu--Se= 2.43, 
			&0.11/1.28\\
			&&a=4.18, c=  9.54 (exp. \cite{savelsberg1978ternare})&K--Se= 3.47\\
			\\
			KCuTe&P6$_3$/mmc&a= 4.48, c= 10.31&Cu--Te= 2.59, 
			&0.52/1.62\\	
			&&a= 4.46, c= 9.95 (exp.\cite{savelsberg1978ternare,dahliah2021high})&K--Te= 3.65\\
		\end{tabular}}}
  
	\end{ruledtabular}
\end{table*}

\subsection{Electronic Structure}
All the ternary ACuX compounds were theoretically optimized in their respective stable structures to calculate the electronic properties. The prototype crystal structures are shown in Fig~\ref{fig1}, and the crystallographic parameters, such as lattice constants and bond lengths are tabulated in Table~\ref{tab:table2}. The optimized structural details are in fair agreement with the available experimental \cite{savelsberg1978ternare,savelsberg1978darstellung} as well as theoretically \cite{dahliah2021high,parveen2018exploring,boualleg2022ab} reported data. For ACuSe/Te which stabilizes in the hexagonal structure, Cu and Se/Te ions form honeycomb (hexagonal) layers along the c-direction, and between each hexagonal bilayer, there exists a triangular lattice of Na/K atoms (see Fig.~\ref{fig1}(a)). The Na/K atoms are situated on top of the center of the honeycomb lattice, while Cu/Se/Te atoms are positioned on top of the center of the triangular lattice. In contrast, the lattice arrangement in orthorhombic structures for both the space groups (for Pna2$_1$ and Pnma) is slightly different, as shown in Fig~\ref{fig1}(b and c). In both the cases, the X-Cu-X shows a linear chain with an angle of 180$^{\circ}$. 

\begin{figure*} [t!]
	\centering
	\includegraphics[scale=0.52]{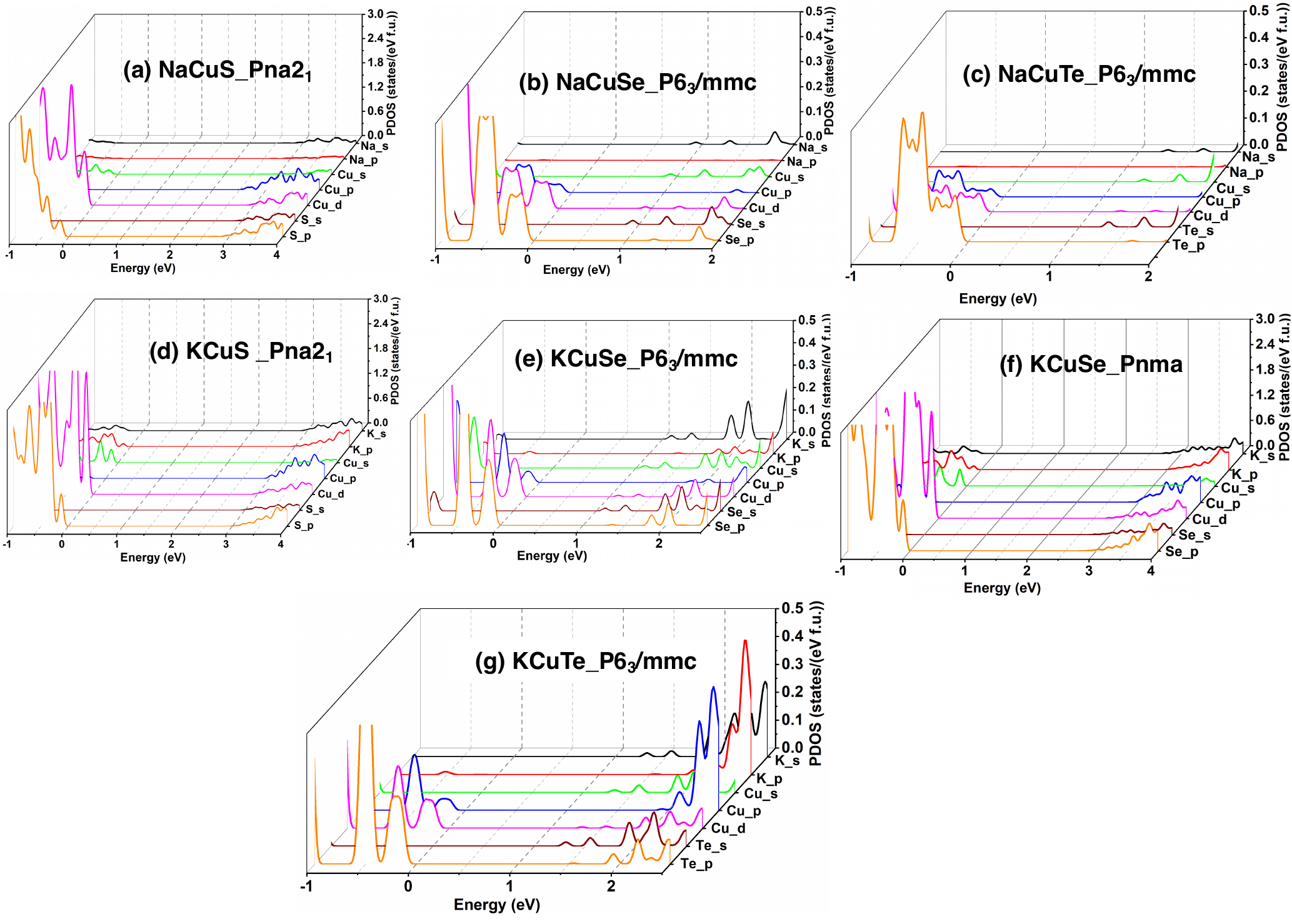}
	\caption{Orbital projected partial density of states (PDOS) for ACuX (A=Na,K; X=S, Se, Te) compounds in the same respective structures as shown in Fig. \ref{fig2}. E$_\mathrm{f}$ is set at 0 eV.}
	\label{fig3}
\end{figure*}

The comparative band structure plots with and without spin--orbit coupling (SOC) within the PBE level are shown in Fig.~S3 of SI\cite{supplment}. As evident from Fig.~S3, there is no change in band gap as well as band dispersion for S- and Se-based compounds. However, Te, a heavy element, slightly changes the band gap apart from the minor band splitting due to the SOC effect. For instance, E$_\mathrm{g}$ for NaCuTe changes from 0.49 eV (w/o SOC) to  0.36 eV (with SOC), and that for KCuTe changes from 0.52 eV (w/o SOC) to 0.41 eV (with SOC) in P6$_3$/mmc structure. Our PBE-SOC results for KCuSe and KCuTe corroborate well with other theoretical reports\cite{parveen2018exploring}. As the SOC effect is not too significant, yet the band gap is conventionally underestimated within the PBE functional, we further performed the electronic structures of all the compounds using HSE06 functional without the SOC effect for more accurate band gap estimation.

Figure~\ref{fig2} shows the electronic band structures and the optical transition probability of all the ACuX compounds using hybrid (HSE06) functionals. All the compounds are direct band gap semiconductors in which the valance band maximum (VBM) and conduction band minimum (CBM) are located at $\Gamma$--point. Table~\ref{tab:table2} display the band gap (E$_\mathrm{g}$) values simulated using both PBE and HSE06 functionals. We found that the E$_\mathrm{g}$ lies in the range 0.83 eV to 2.88 eV for different compounds. Clearly, E$_\mathrm{g}$ for NaCuSe (1.21 eV), KCuSe (1.28 eV), and KCuTe (1.62 eV) are most suitable for photovoltaic applications. It is also observed that the VBM at the $\Gamma$-point is doubly degenerate for crystal structures belonging to hexagonal
symmetry \emph{i.e.,} P6$_3$/mmc of ACu(Se/Te) (Fig.~\ref{fig2}(b,c,e,g)). A flat valence band edge is also observed in all the compounds, which can be very helpful for promising carrier transport due to their high effective masses.

Figure~\ref{fig3} shows the orbital projected partial density of states (PDOS) for all the ACuX compounds in the respective structures shown in Fig. \ref{fig1}. The PDOS reveals that the VBM mainly consists of the p- and d-orbitals of Cu and the p- orbital of S/Se/Te atoms at/near E$_\mathrm{f}$ for all the cases. The CBM is mainly contributed by the s-orbital of Na/K,  p- and d-orbitals of Cu, and the s- and p-orbitals of S/Se/Te, respectively. It is also observed that the electronic states near E$_\mathrm{f}$ for alkali metals (Na or K) show a minimal contribution at VBM. The contributions of s-orbitals of alkali metals decrease near E$_\mathrm{f}$  as we move from ACuS to ACuSe and ACuTe. Unlike several other chalcogenide based compounds, the E$_\mathrm{g}$ value in the present case first decreases as we go from Na/KCuS to Na/KCuSe, and then increases for Na/KCuTe. For example, E$_\mathrm{g}$ for NaCuS, NaCuSe and NaCuTe are 2.42 eV, 0.83 eV, and 1.21 eV respectively and a similar trend is obtained for KCuX systems as well. This is mainly attributed to the nature of hybridization between Cu and chalcogen atoms. The electronegativity values for differnt atoms are $\chi_{Na}$= 0.93, $\chi_{K}$= 0.82, $\chi_{Cu}$= 1.9, and $\chi_{S/Se/Te}$= 2.58/2.55/2.1, respectively. As chalcogen atoms are more electronegative than alkali metals, there is always an electronic charge transfer from the Na/K atom to S/Se/Te atoms. Similarly, the charge is also transformed from Cu atoms to S/Se atoms. However, a negligibly small charge transfer between Cu and Te atoms can be expected because their electronegativity is almost similar (the difference is 0.2). This is also reflected in their orbital PDOS plots. 
In order to better understand the role of hybridization in dictating the E$_\mathrm{g}$ trend, we show a zoomed-in view of the orbital projected PDOS above CBM for  NaCuS, NaCuSe, and NaCuTe, respectively in {Fig.~S4(a,b,c) of SI\cite{supplment}. A careful inspection of these plots clearly differentiate the orbital hybridization between Cu and S/S/Te atoms at the CBM side. The electronic states of the p- and d- orbitals of Cu and the p orbitals of S atoms mainly contribute to the CBM side of NaCuS. Similarly, the CBM of NaCuSe consists of the s- and d-orbitals of Cu and the s-orbital of Se. However, the nature of orbital hybridization in NaCuTe is quite different as compared to NaCuS and NaCuSe. Due to minimal charge transfer between Cu and Te, the orbital contributions from Cu--p, Cu--d, and Te--s are diminished near E$_\mathrm{f}$ and indicate only a strong hybridization between Na--s, Cu--s, and Te--p at the CBM side of NaCuTe, causing an increase in the band gap. Similar band gap trend and the orbital hybridization are also observed between Cu and Te atoms in KCuX compounds (see Fig.~S4(d,e,f) of SI \cite{supplment}) and can be explained on a similar ground. An analogous band gap trend is also observed in other chalcogenide based compounds like CdS, CdSe, and CdTe as well as ZnS, ZnSe, and ZnTe, respectively\cite{deligoz2006elastic,duan1998electronic,zakharov1994quasiparticle}. 

\begin{figure*}[htbp]
	\centering
	\includegraphics[scale=0.67]{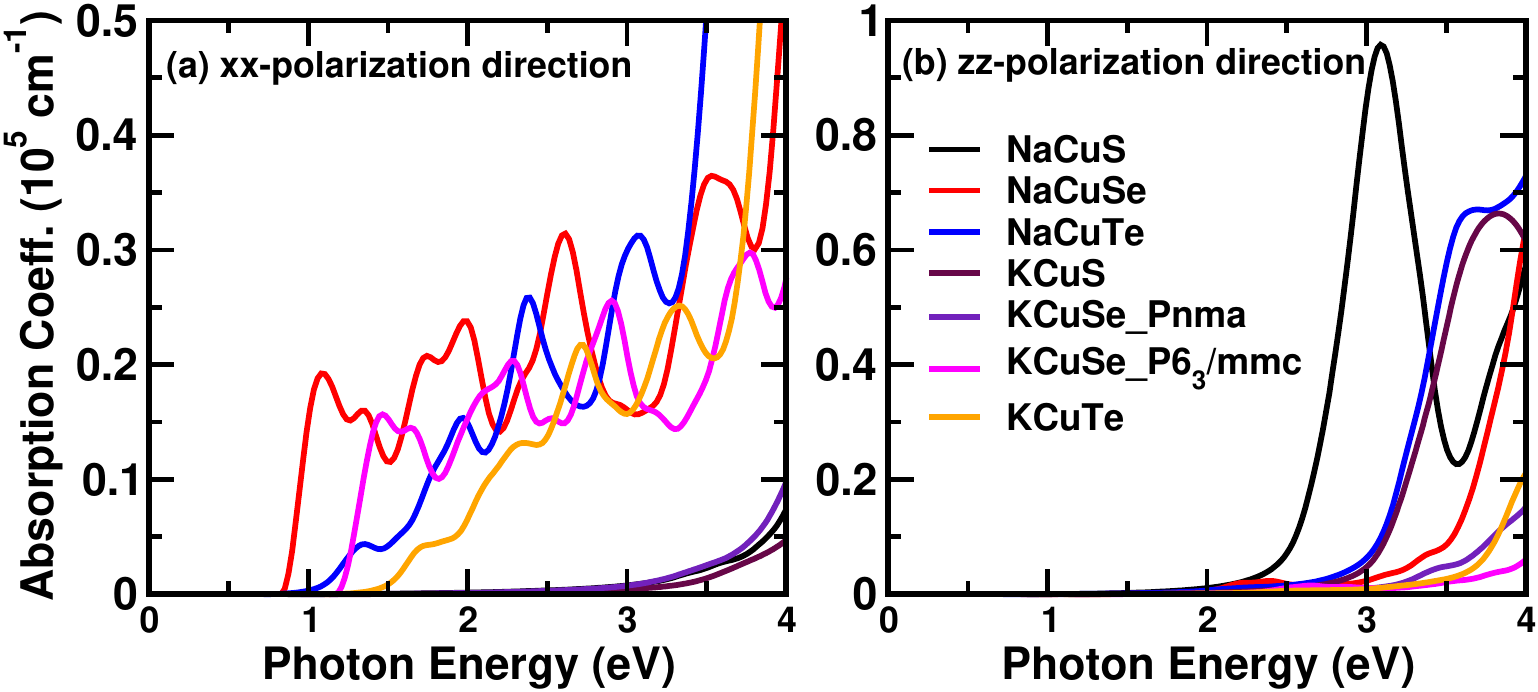}
	\caption{Absorption coefficient ($\alpha$) for ACuX (A=Na,K; X=S, Se, Te) along (a) x-- and (b) z--polarization directions.}
 
	\label{fig4}
\end{figure*}

 \subsection{Optical Properties}
The band structure calculations established that all the ACuX compounds are direct band gap semiconductors, with the band gap falling in the range 0.83 eV to 2.88 eV. This motivates us to further study the optical properties of these systems and evaluate their efficacy as solar absorber. Along with the absorption coefficient, we have calculated another screening parameter \emph{i.e.,} spectroscopic limited maximum efficiency (SLME) proposed by Yu \emph{et al.} \cite{yu2012identification}. SLME gives an upper bound on the solar efficiency by incorporating the nature/magnitude of band gap and absorption coefficient for a particular compound. It is is an improvised version of the Shockley--Queisser (SQ) efficiency limit. Simulation of SLME also require the information about the possibility of optical transition from VBM to CBM for all the ACuX compounds. To obtain this, we have calculated the transition probability (p$^2$) by calculating the square of transition dipole matrix elements. The transition probability for all the ACuX systems are shown in Fig.~\ref{fig2} (below each band structure plot). Clearly, direct optical transition from VBM to CBM is allowed for all the hexagonal structures of ACuSe/Te due to the finite p$^2$ values at the high--symmetry point $\Gamma$. The optical transitions are allowed due to the same parity transition of electron states from p--states of S/Se/Te atoms at the VBM side to s-- and p-- states of Cu at the CBM side and d--states of Cu at VBM to s--states of S/Se/Te at CBM. In contrast, the p$^2$ value is found to be zero at $\Gamma$--point for  NaCuS, KCuS, and KCuSe (orthorhombic structures) in Fig.~\ref{fig2}(a,d and f), indicating optically forbidden transition from VBM to CBM.

The absorption coefficient ($\alpha$) of a given material is a quantifiable descriptor which dictate the penetration extent of a photon (with a  particular wavelength) into the material before it gets absorbed. For a suitable solar absorber, a sharp rise in the absorption coefficient is obtained after the incidence of photon energy closer to its band gap. $\alpha$ is related to the dielectric function and can be calculated using the following expression:
\begin{equation}
\alpha(E)=\frac{\sqrt{2}\omega}{c}\bigg[\sqrt{\sqrt{\varepsilon_1^2(\omega)+\varepsilon_2^2(\omega))}-\varepsilon_1}\bigg]
\end{equation}

where $E$ is the incident photon energy, $\omega$ is the angular
frequency related to photon energy via $E=\hbar$$\omega$ ($\hbar$ is the reduced Planck’s constant), and $c$ is the speed of light in vacuum, respectively. $\varepsilon_1$ and $\varepsilon_2$ are the real and imaginary parts of the dielectric function.  Figure~\ref{fig4}(a,b) shows the absorption spectra plot for all ACuX compounds along the x--  (left) and z--polarization (right) directions (arising out of structural anisotropy in hexagonal and orthorhombic structures). For orthorhombic systems (NaCuS, KCuS, KCuSe), the absorption spectra along the y--polarization direction are shown in Fig.~S5 of SI\cite{supplment}.  The absorption onset for all the compounds is shifted to the band gap obtained from hybrid HSE06 calculations. Clearly, the absorption coefficients are not same along the three polarization directions because of the associated structural anisotropy. For example, in orthorhombic systems, the absorption coefficient of NaCuS and KCuS is higher ($\sim$ 0.6 to 0.9$\times$10$^5$ cm$^{-1}$) along the z--polarization direction than the x-- and y--direction. However, the absorption coefficient of orthorhombic KCuS is high along y--polarization direction with respect to x-- and z--direction (see Fig.~S5). Similarly, for the hexagonal structures \emph{i.e.,} KCuSe, KCuTe, NaCuSe, and
NaCuTe, the simulated absorption coefficients fall in the range 0.2$\times$10$^5$ to 0.6$\times$10$^5$ cm$^{-1}$ along x-- and y--polarization directions, with a relatively small contribution arising from the z--direction  as shown in Fig.~\ref{fig4}(b).

\begin{figure*}[htbp]
	\centering
	\includegraphics[scale=0.55]{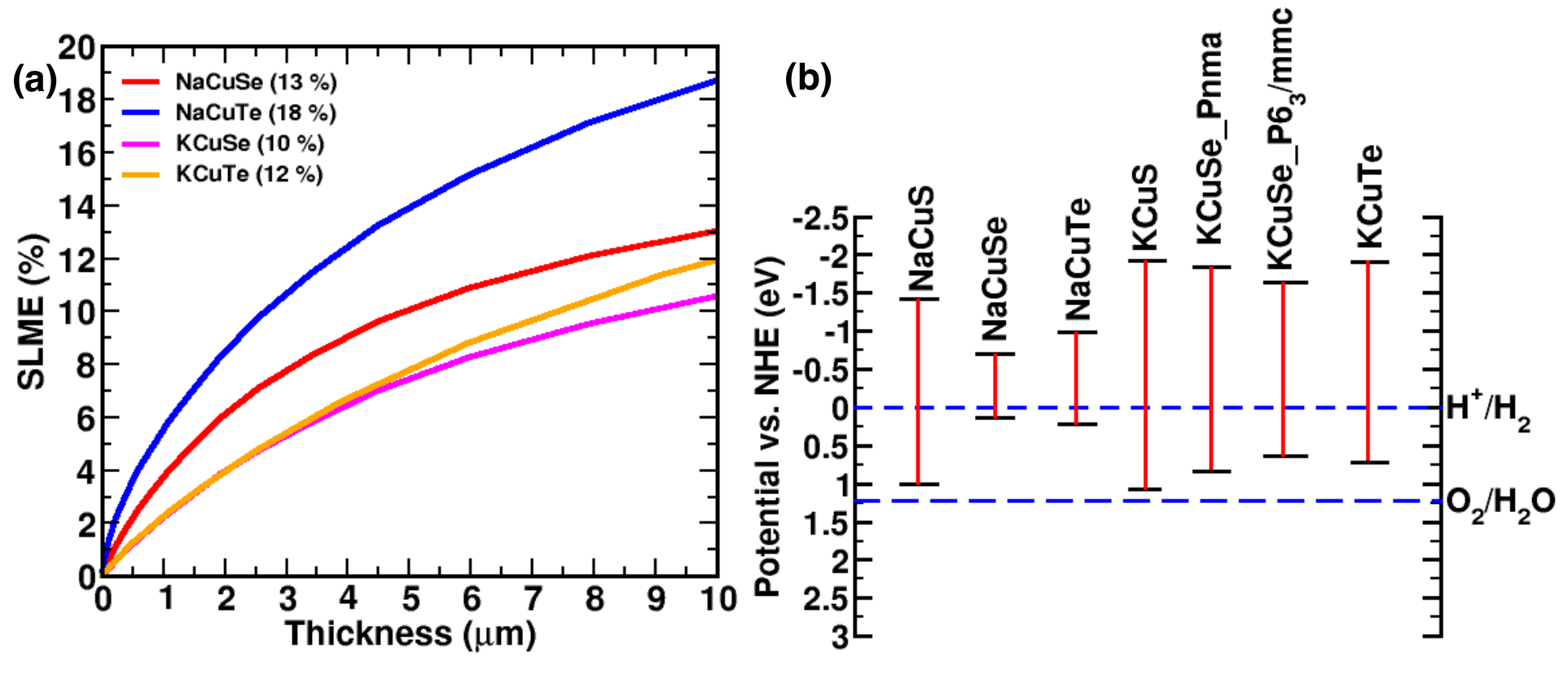}
	\caption{(a) Spectroscopic Limited Maximum Efficiency, SLME (at 300 K) vs. film thickness for the hexagonal ACuX compounds whose band gaps lie in the visible region. (b) Band edge positions with respect to water redox levels for all stable ACuX compounds. }
	\label{fig5}
\end{figure*}

 Next, we have calculated the thickness dependence of SLME of all the ACuX compounds, considering the application for thin film solar cells. We have chosen four compounds \emph{i.e.,} hexagonal ACu(Se/Te) (A=Na, K) because of their allowed optical transition and high absorption coefficients in the visible region. The thickness dependence of SLME is shown in Fig~\ref{fig5}(a). Clearly, with increasing thickness, the SLME increases and then saturates beyond a certain thickness. The maximum SLME is found to be 18 \% for a 10 $\mu$m thick NaCuTe film. Similarly, for the rest of the ACuX systems, the SLME lie between $\sim$10 to 13\%. Based on the optical and SLME descriptors, it is reasonable to consider that all the selenide and telluride based ACuX systems can be potential candidates for light absorbers in solar PV devices.

\subsection{Potential Application in Photoelectrochemical (PEC) Water Splitting}
We further investigate the feasibility of the ternary ACuX compounds for photocatalytic water splitting applicaions. We calculated the band edge positions with respect to the water redox levels\cite{xu2000absolute,castelli2015new} using the empirical model proposed by Butler and Ginley\cite{butler1978prediction}. According
to their model, the valence and conduction band edge positions can be obtained from the following expression,
\begin{equation}
E_{VBE/CBE} = E_0+(\chi_A\chi_{Cu}\chi_X)^{1/3}\pm \frac{E_g}{2}
\end{equation}
where $E_0$ is the difference between a normal hydrogen electrode (NHE) and vacuum whose value is -4.5 eV. $\chi$ is the electronegativities of the constituent elements in the Mulliken scale, and $E_\mathrm{g}$ is the band gap. Figure~\ref{fig5}(b) shows the band edge positions for all the stable ACuX compounds. For a compound to be promising for PEC water splitting, the CBM must locate more negative than the redox potential of H$^+$/H$_2$ (0 V \emph{vs.} NHE), and the VBM must align more positive than the redox potential of O$_2$/H$_2$O (1.23 V \emph{vs.} NHE) at ambient condition\cite{chakrapani2007charge}. As evident from Fig.~\ref{fig5}(b), it is clear that all ternary copper chalcogenides have a well positioned conduction band edge to be used as photocathode for hydron evolution reaction (HER) in a PEC cell. 

\section{Conclusion}
In conclusion, we have systematically investigated the chemical/dynamical stability and optoelectronic properties of ternary ACuX (A=Na, K; X=S, Se, Te) chalcogenides using first--principles calculations. The structural stability metrics ensure that NaCuS, KCuS and KCuSe crystallize in orthorhombic structure while NaCuSe/Te and KCuTe stabilize in hexagonal phase. Interestingly, few of these compounds show intriguing energetics/phonon trends, which suggests them to stabilize in structure(s) other than that predicted experimentally (one report only). This demands for a revisit of the crystal structure prediction. All the compounds are direct band gap semiconductors having band gaps laying between 0.83 eV to 2.88 eV. The direct optical transitions are forbidden for orthorhombic systems (NaCuS, KCuS, and KCuSe), while it is allowed for the remaining systems (hexagonal phase). The absorption coefficient for the optically allowed ACuX compounds lie in the range $\sim$(0.2 to 0.6)$\times$10$^5$ cm$^{-1}$ in the visible region. The highest simulated efficiency for the NaCuTe is determined to be 18\% for an 8 $\mu$m thick film.  In addition, the favorable straddling of valence band edges with respect to the normal hydrogen electrodes confirms the suitability of all these compounds as photocathodes for the hydrogen evolution reaction (HER) in the photoelectrochemical (PEC) process. The present study suggests that the ternary alkali metal-based copper chalcogenides can be suitable candidates as light absorbers in PV cells as well as photocathodes for HER in photocatalytic water splitting applications, requiring further experimental investigation.

\section*{Acknowledgements}
G.B. would like to thank the Council of Scientific and Industrial Research (CSIR), India, for providing senior research fellowship. N.S. acknowledge the financial support from the Khalifa University of Science and Technology under the Emerging Science \& Innovation Grant ESIG-2023-004 and the contribution of Khalifa University's high-performance computing and
research computing facilities to the results of this research.

\bibliography{behera_paper.bib}

\end{document}